\documentclass[aps,prl,twocolumn,superscriptaddress,noshowpacs]{revtex4-2}

\usepackage{eqnarray}
\usepackage{amsmath,amsfonts,amssymb}

\usepackage{graphicx}
\usepackage{bm}
\usepackage{color}
\usepackage{physics}

\begin{document}

\title{Kibble-Zurek mechanism in a polariton supersolid}

\author{Dmitry Solnyshkov}
\email{dmitry.solnyshkov@uca.fr}
\affiliation{Institut Pascal, PHOTON-N2, Universit\'e Clermont Auvergne, CNRS, Clermont INP,  F-63000 Clermont-Ferrand, France}
\affiliation{Institut Universitaire de France (IUF), 75231 Paris, France}

\author{Rafal Mirek}
\affiliation{IBM Research Europe - Zurich, Säumerstrasse 4, Rüschlikon, Switzerland}

\author{Darius Urbonas}
\affiliation{IBM Research Europe - Zurich, Säumerstrasse 4, Rüschlikon, Switzerland}

\author{Etsuki Kobiyama}
\affiliation{IBM Research Europe - Zurich, Säumerstrasse 4, Rüschlikon, Switzerland}

\author{Pietro Tassan}
\affiliation{IBM Research Europe - Zurich, Säumerstrasse 4, Rüschlikon, Switzerland}

\author{Ioannis Georgakilas}
\affiliation{IBM Research Europe - Zurich, Säumerstrasse 4, Rüschlikon, Switzerland}

\author{Rainer F. Mahrt}
\affiliation{IBM Research Europe - Zurich, Säumerstrasse 4, Rüschlikon, Switzerland}

\author{Michael Forster}
\affiliation{Macromolecular Chemistry Group and Wuppertal Center for Smart Materials \& Systems (CM@S), Bergische Universit\"{a}t, Wuppertal, Gauss Strasse 20, 42119 Wuppertal, Germany}
\author{Ullrich Scherf}
\affiliation{Macromolecular Chemistry Group and Wuppertal Center for Smart Materials \& Systems (CM@S), Bergische Universit\"{a}t, Wuppertal, Gauss Strasse 20, 42119 Wuppertal, Germany}

\author{Marcin Muszynski}
\affiliation{Institute of Experimental Physics, Faculty of Physics, University of Warsaw, Warsaw, Poland}

\author{Wiktor Piecek}
\affiliation{Institute of Applied Physics, Military University of Technology, Warsaw, Poland}

\author{Piotr Kapu\'sci\'nski}
\affiliation{Institute of Experimental Physics, Faculty of Physics, University of Warsaw, Warsaw, Poland}
\author{Jacek Szczytko}
\affiliation{Institute of Experimental Physics, Faculty of Physics, University of Warsaw, Warsaw, Poland}

\author{Thilo Stoferle}
\affiliation{IBM Research Europe - Zurich, Säumerstrasse 4, Rüschlikon, Switzerland}

\author{Guillaume Malpuech}
\email{guillaume.malpuech@uca.fr}
\affiliation{Institut Pascal, PHOTON-N2, Universit\'e Clermont Auvergne, CNRS, Clermont INP,  F-63000 Clermont-Ferrand, France}

\begin{abstract}
We study the formation of topological defects via the Kibble-Zurek mechanism in a polariton supersolid in a liquid crystal microcavity with tunable Rashba-Dresselhaus spin-orbit coupling. We predict analytically two different scalings in the slow- and fast-quench regimes, and confirm these predictions numerically. We also present experimental results for the slow-quench regime, demonstrating an original Kibble-Zurek scaling exponent $\eta_{KZM}=1.0\pm 0.2$
\end{abstract}

\maketitle

Supersolidity is currently an active topic of research. After the initial suggestion~\cite{leggett1970can} and years of research in different platforms, significant progress has been achieved recently: supersolidity has been reported 
in different types of atomic condensates~\cite{leonard2017supersolid,li2017stripe,geier2023dynamics, tanzi2021evidence,biagioni2024measurement,casotti2024observation} and also claimed in various optical systems~\cite{trypogeorgos2025emerging,muszynski2024observation,zhai2025electricallytunablenonrigidmoire}. One of the important mechanisms responsible for the density modulation associated with a supersolid is the spin-orbit coupling~\cite{koralek2009emergence,lin2011spin}, capable of modifying the dispersion and creating two degenerate states for condensation. The interference of these states leads to the periodic density pattern (stripes) associated with supersolidity. The interference is not the result of collision of two propagating condensates: it appears from two spatially coexisting motionless condensates. The other characteristic feature of a supersolid, beyond the spatial order, is the superfluidity. It appears naturally in a condensate with weak repulsive interactions, if it is not flowing faster than the critical velocity.

\begin{figure}
    \centering
    \includegraphics[width=\linewidth]{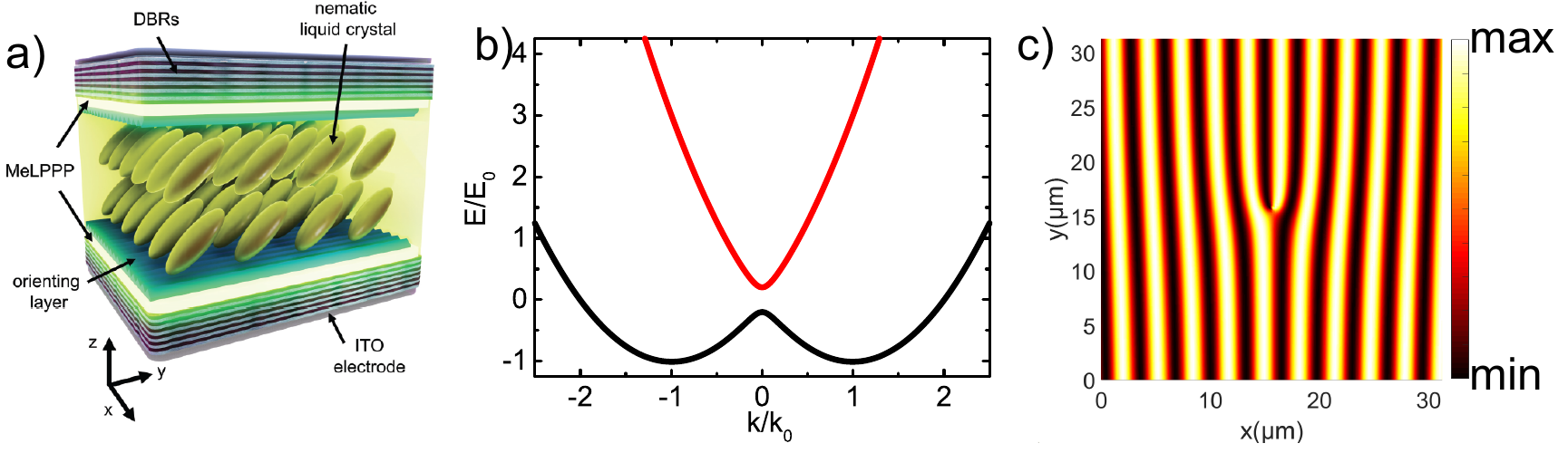}
    \caption{Polariton supersolid in a liquid crystal microcavity. a) The microcavity with a polymer and a liquid crystal orientable by electric field. b) The lower polariton branch with Rashba-Dresselhaus spin-orbit coupling (black and red are the two subbands). c) The periodic density profile of a supersolid with a topological defect (quantum vortex) that shows up as a dislocation.}
    \label{fig1}
\end{figure}

Recently, a particularly versatile photonic platform has been suggested as a powerful tool for fundamental and applied studies based on Hamiltonian engineering~\cite{rechcinska2019engineering}. A planar microcavity filled with a liquid crystal providing tunable birefringence (Fig.~\ref{fig1}) allows putting two orthogonally-polarized modes of different parity into resonance, giving rise to the so-called Rashba-Dresselhaus spin-orbit coupling with equal strength (RDSOC or emergent optical activity~\cite{ren2021nontrivial}). Photonic RDSOC has already demonstrated a rich phenomenology~\cite{krol2021observation,lempicka2022electrically,li2022manipulating}.
This coupling is linear in wave vector, and thus modifies the dispersion relation, creating two degenerate energy minima.
Inserting a material with strong excitonic resonances is such cavities allows achieving RDSOC for exciton-polaritons \cite{lempicka2022electrically,liang2024polariton,muszynski2024observation} which are  interacting photons \cite{kavokin2017microcavities,carusotto2013quantum} able to undergo BEC effects. A polariton supersolid has been indeed reported \cite {muszynski2024observation} by using as an active media a polymer layer (MeLPPP~\cite{scherf1992melppp})  with a strong excitonic resonance, which ensures thermalization, condensation and weak interactions of exciton-polaritons via light-matter coupling at room temperature~\cite{plumhof2014room,zasedatelev2021single}.

Among other aspects, supersolidity is promising because it could explain the glitches of neutron stars (sudden jumps in their rotation frequency), linked with the quantization of the rotation of the supersolids via the vortices~\cite{Poli2023}. These quantum vortices were observed in atomic~\cite{casotti2024observation} and polariton~\cite{muszynski2024observation} supersolids only very recently.
One of the particularly interesting phenomena associated with topological defects such as quantum vortices is the Kibble-Zurek mechanism~\cite{Kibble1976,Zurek1985,Zurek1996} (KZM), responsible for their formation during phase transitions such as U(1) symmetry breaking.
The spontaneous formation of a supersolid involves spontaneous breaking of two U(1) symmetries: one of them corresponds to the choice of the phase of the condensate wavefunction, and the other -- to the choice of the phase determining the spatial position of the stripes. If a supersolid is implemented in a spinor condensate, these two phases can be seen as either the average (global phase) and the relative phase of the two components or as the individual phases of each of the two components. 
While the KZM in general has been studied a lot in various systems, including atomic and polariton~\cite{Damski2010,Chen2011quench,Mat2014,Anquez2016,Autti2016,Solnyshkov2016kzm,Mat2017phase,Comaron2018,GKZM,KKZM} condensates of different dimensionality and also in presence of external lattices, the research on KZM in supersolids is only starting~\cite{Kirkby2025}. In particular, there have been no theoretical predictions and no experiments studying the KZM during supersolid formation via the spin-orbit coupling.

In this work, we present a theoretical and experimental study of the Kibble-Zurek mechanism of formation of quantum vortices during the condensation of polaritons in a supersolid phase in presence of Rashba-Dresselhaus spin-orbit coupling in a tunable liquid crystal microcavity. We analyse the structure of individual vortices in polariton supersolids. We  show how the modification of the dispersion relation by the RDSOC affects the scaling of the density of topological defects appearing during the quench, with two different characteristic exponents found in the slow and fast quench regimes. These two exponents (1/3 and 1) differ from the one expected (1/2) for a 2D spinor superfluid with spin anisotropic interactions (in the absence of SOC). Finally, we present the results of an experiment with $10^4$ single-shot images verifying our theoretical predictions.

\textit{The condensate.}
The Hamiltonian of the bare polariton modes on the circular polarization basis (neglecting TE-TM, parabolic approximation) under the coupling of two orthogonally-polarized modes with opposite parity~\cite{rechcinska2019engineering,ren2021nontrivial} is given by
\begin{equation}
H = \left( {\begin{array}{*{20}{c}}
{\frac{{{\hbar ^2}{k^2}}}{{2m}} + \alpha k_x}&{\Delta  }\\
{\Delta  }&{\frac{{{\hbar ^2}{k^2}}}{{2m}} - \alpha k_x}
\end{array}} \right)
\end{equation}
where $m$ is the polariton mass, $\alpha$ is the RDSOC constant, and $\Delta$ is the so-called detuning between the modes of orthogonal polarization (residual birefringence).
An example of the resulting dispersion along $k_x$ of a pair of coupled polariton modes is shown in Fig.~\ref{fig1}(b) demonstrating two minima at $k_x=k_0$. Note that the dispersion along $k_y$ is simply parabolic. Modes in the minima are elliptically polarized with an opposite circular polarization degree but the same linear polarization degree which is on the order of 10 \% in typical experimental conditions. 

Non-resonant pumping injects excitons into a reservoir at high energies. These excitons relax and form polaritons, which, if the density is sufficient, undergo Bose-Einstein condensation~\cite{kasprzak2006bose} in the two minima of the dispersion~\cite{muszynski2024observation}. Assuming perfect thermal equilibirum at $T=0$~K, it corresponds to the formation of a four-component condensate with two spins and two wave vectors whose wave function, on a circular polarization basis reads:
\begin{equation}
\left| \psi  \right\rangle  =\frac{1}{2} \left( {\begin{array}{*{20}{c}}
{\cos \frac{\theta }{2}}\\[2pt] 
{\sin \frac{\theta }{2}}
\end{array}} \right){e^{ - ik_0x-i\phi_1}} + \frac{1}{2}\left( {\begin{array}{*{20}{c}}
{\cos \frac{{\pi  - \theta }}{2}}\\[2pt] 
{\sin \frac{{\pi  - \theta }}{2}}
\end{array}} \right){e^{ + ik_0x-i\phi_2}}
\label{WF1}
\end{equation}
where $\theta$ is the polar angle of the Stokes vector. The phases $\phi_1$ and $\phi_2$ of the two k-components are chosen randomly due to the spontaneous symmetry breaking during the condensation process. 
The expression can be rewritten as
\begin{equation}
\left| \psi  \right\rangle  =\frac{1}{2}e^{i\zeta}\left( \left( {\begin{array}{*{20}{c}}
{\cos \frac{\theta }{2}}\\[2pt] 
{\sin \frac{\theta }{2}}
\end{array}} \right){e^{ - i(k_0x-\eta)}} + \left( {\begin{array}{*{20}{c}}
{\cos \frac{{\pi  - \theta }}{2}}\\[2pt] 
{\sin \frac{{\pi  - \theta }}{2}}
\end{array}} \right){e^{ + i(k_0x-\eta)}}\right)
\label{WF2}
\end{equation}
where the global phase is $\zeta=-(\phi_1+\phi_2)/2$, and the relative phase controlling the position of the stripes is $\eta=-(\phi_1-\phi_2)/2$ (see \cite{suppl} for more details).
The total density shows stripes with a contrast equal to the linear polarization degree $\sin {\theta}$ in the two minima of the dispersion. The linear polarization degree shows stripes with a contrast of 1, whereas the circular polarization should not show stripes. Fig.~\ref{fig1}(c) shows a typical stripe pattern but also a dislocation in the middle of the figure, which is induced by a quantum vortex present in the wave function of one of the wave vector components of the condensate.

\textit{Types of vortices}
When $\Delta=0$, there is a spin-valley locking ($\theta=0$)  and spin components are uniquely attributed to a given wave vector $\pm k_0$. Elementary vortices are half-vortices present only in one spin-valley component. When $\Delta$ is nonzero, both spin components are present in both k-valleys but a vortex should be present for the two spin components of a given valley in order to preserve the linear polarization orientation of the k-state far from the core (see~\cite{suppl}). A given valley cannot host a vortex of winding 1 for one spin component and 0 or -1 for the other spin component, for instance.
So, only four types of single-charge vortices exist with winding $\pm 1$ in two valleys ($k_0$ or $-k_0$) respectively. These vortices appear as fork-like dislocations both in the total density and linear polarization degree (see~\cite{suppl}) and are directly detected by our setup.

\textit{Kibble-Zurek mechanism} The Kibble-Zurek mechanism~\cite{Zurek1985} of the formation of the topological defects arises from the impossibility of crossing any second-order phase transition point adiabatically, because of the divergence of the relaxation time at the transition point. The fluctuations forming above the transition point become frozen, and their size determines the distance between the resulting topological defects.
The scaling of the density of these  defects versus the quench parameter $\varepsilon$ is defined by two scaling exponents: $\nu$, the critical exponent of the correlation length $\xi$ , defined by
\begin{equation}
\xi \left( \varepsilon  \right) = \frac{{{\xi _0}}}{{{{\left| \varepsilon  \right|}^\nu }}}
\end{equation}
and the dynamic critical exponent $z$ appearing in the relaxation time $\tau$ as
\begin{equation}
    \tau(\varepsilon)=\frac{\tau_0}{|\varepsilon|^{z\nu}}
\end{equation}
In polariton condensation, the temperature often remains constant~\cite{kasprzak2006bose}, while the quench parameter is the normalized pumping density which triggers the condensate buildup from an empty system.

\textit{Experimental results.} 
We have performed pulsed excitation and measured polarization dependent spatially resolved intensity integrated in time over each single-shot excitation as previously reported in ~\cite{muszynski2024observation}. The sample used is among those studied in ~\cite{muszynski2024observation} and its characteristics are detailed in the Supplementary~\cite{suppl}. The power of the pumping laser was slowly varying close to the condensation threshold so that the condensate density was also varying over one order of magnitude. This has allowed us to acquire a sufficient number of single-shot images to study the statistics of the vortices densities versus pumping power (about 40 shots for each intensity, on average).

\begin{figure}
    \centering
    \includegraphics[width=0.95\linewidth]{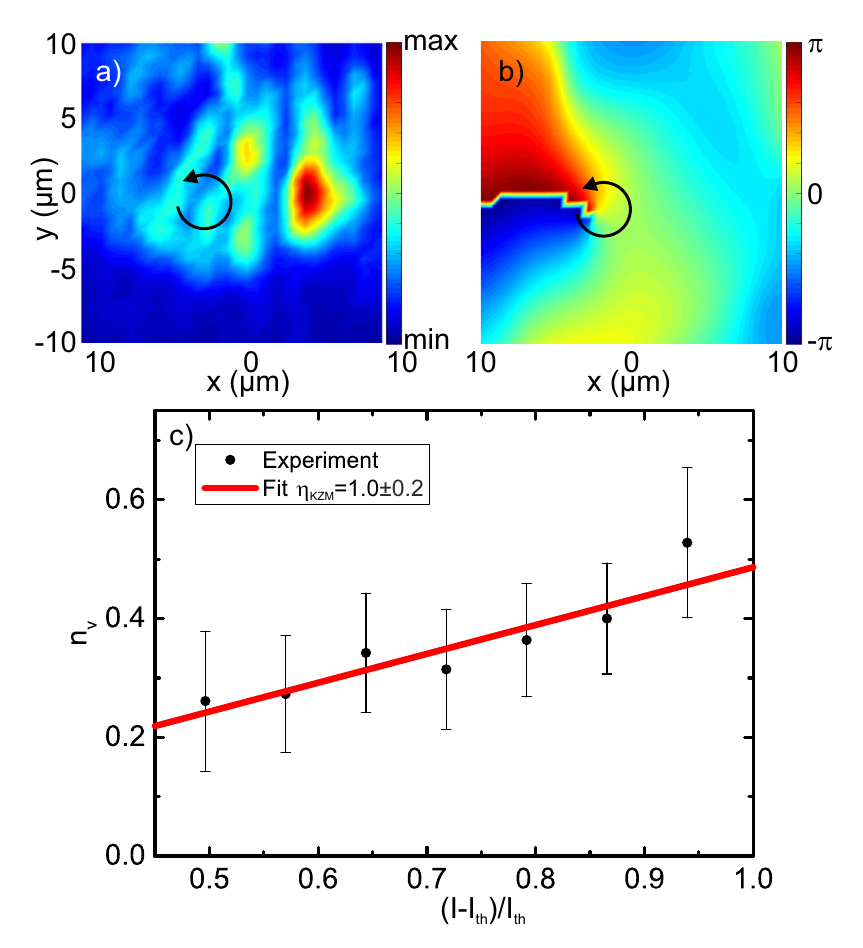}
    \caption{Experimental KZM scaling in a supersolid.
    a) Intensity of emission from a supersolid, demonstrating the stripes with a dislocation.  b) Phase, extracted from the pattern of the stripes, demonstrating a singularity (quantum vortex). Black arrow shows the phase gradient in (a,b). c) Average number of vortices in a single shot as a function of reduced intensity. Black dots -- experiment, red line -- theory. Error bars correspond to the uncertainty provided by the Poissonian distribution fit.}
    \label{fig3}
\end{figure}

The results of experiments are shown in Fig.~\ref{fig3}. An example of a single-shot image with density stripes demonstrating a dislocation is shown in Fig.~\ref{fig3}(a). The extracted phase of the stripes is shown in Fig.~\ref{fig3}(b). The vortex appears as a singularity, the direction of the gradient of the phase is shown with a black arrow. For fixed pumping, and in the central region where the condensate density is large and approximately uniform, the positions of the vortices are entirely random from shot to shot (see Fig.~S1(a) in~\cite{suppl}), with the statistics of their position well described by the Poisson distribution (see Fig.~S1(b) in~\cite{suppl}). The randomness of their positions is a clear signature of the absence of their pinning by disorder, a strong argument in favor of the superfluid character of the condensate~\cite{muszynski2024observation}. 

A plot of the number of vortices as a function of the intensity relative to the threshold (quench parameter) is shown in Fig.~\ref{fig3}(c) with black dots. The number of vortices is extracted from the phase distribution for each single shot automatically. For each intensity, the data on the number of vortices in each shot is fitted with the Poisson distribution, allowing to obtain the average number of vortices and the uncertainty (used to plot the error bars). The red line represents a power law fit giving a scaling exponent $\eta_{KZM,exp}=1.0\pm 0.2$. To explain this result, we now focus on the theoretical description of the supersolid formation in polariton condensate with RDSOC.

\textit{Theory.}
To describe the polariton relaxation and the formation of the condensate, we use the spinor Gross-Pitaevskii equation including spin-anisotropic polariton-polariton interactions
\begin{eqnarray}
i\hbar \frac{{\partial {\psi _ \pm }}}{{\partial t}} &=& \left( {1 - i\Lambda } \right){\hat T_ \pm }{\psi _ \pm }  + g_1 |\psi_{\pm}|^2\psi_{\pm}\nonumber\\
&+&g_2 |\psi_{\mp}|^2\psi_{\pm}+i\gamma(|\psi|^2){\psi _ \pm }\nonumber\\
&+&\Delta\psi_\mp+\chi.
\label{gpe}
\end{eqnarray}
Here, $\hat{T}_\pm$ is the kinetic energy operator containing terms accounting for several effects: the dispersion relation of polaritons with mass $m$, the gauge potential contribution of the RDSOC $-2i\alpha \cdot \partial\psi_\pm/\partial x$, and the energy relaxation mechanisms~\cite{Pitaevskii58} via $\Lambda$, with zero energy level set at the bottom of the dispersion.  $\gamma(|\psi|^2)$ is the term combining decay and saturated gain, $\chi$ is the noise describing the spontaneous scattering from the excitonic reservoir. The saturated gain is given by $\gamma(\left|\psi\right|^2)=\gamma_0(n_R)\exp(-n_{tot}/n_s)$
with $n_{tot}=\int|\psi|^2\,dxdy$ the total particle density, $n_s$ the saturation density, and $n_R$ the exciton reservoir density, approximately constant at the time scale of the condensate formation thanks to sufficiently long optical pulses~\cite{suppl}. $g_1$ and $g_2$ are the polariton-polariton interaction constants (same-spin and opposite-spin, respectively, with $g_2<0$ and $|g_2|\ll g_1$ for negative detunings~\cite{Vladimirova2010}). 
The pumping power, which is the quench parameter that we vary, is described by the gain $(\gamma_0-\gamma_{th})/\gamma_{th}$, where $\gamma_{th}$ is the threshold value of gain, below which the condensation does not occur. 

In our case, as it happens most often, the energy relaxation means that the losses are directly proportional to the energy: $\Gamma(E)\sim E$ \cite{Pitaevskii58}, which gives $z\nu=1$ for the dynamical exponent. The behavior of the critical exponent of the correlation length is obtained by analyzing the contributions to the energy of the system. Since in our case the energy has a complicated expression with multiple contributions~\cite{suppl}, we can expect the scaling to change depending on the regime. In particular, one can expect a transition to occur when the characteristic wave vector $k^*=1/\xi$ becomes comparable with $k_0$, or, in other words, when the characteristic correlation length $\xi$ becomes comparable to the period of the supersolid $a=\pi/k_0$.

The expected scaling of the density of topological defects can be calculated as
\begin{equation}
    n=\left(\frac{\tau_0}{\tau_q}\right)^{(D-d)\frac{\nu}{1+z\nu}}
\end{equation}
where $D=2$ is the dimensionality of the system, $d=0$ is the dimensionality of the defect (quantum vortex), $\tau_q=\Lambda/(2\Gamma_0 \gamma)$ is the quench time. This expression defines the Kibble-Zurek scaling exponent $\eta_{KZM}=(D-d)\nu/(1+z\nu)$.

First of all, if there is no coupling between the two polarizations ($\Delta=0$), the system can be considered as two independent scalar Gross-Pitaevskii equations with parabolic dispersion (in spite of the $\alpha k$ term, which simply shifts the parabola in this case). Writing $g_1n=\hbar^2 {k^*}^2/(2m)$ allows finding the critical exponent $\nu=1/2$, and therefore $\eta_{KZM,scalar}=1/2$. One expects therefore to observe a square root scaling for zero detuning, without any transitions in this case. This is indeed confirmed by numerical simulations based on~\eqref{gpe} (not shown).

Let us now consider a non-zero detuning $\Delta$, starting with the case $\xi\gg a$, which corresponds to small wave vectors (large characteristic distance between the topological defects, slow quench regime). In this case, the linear contribution to the kinetic energy arising from the RDSOC dominates, and one can write the equality determining the scaling of the correlation length as $\alpha k^* = g_1n $ which gives $\xi=\alpha/|g_1n|$. The corresponding scaling exponent is $\nu=1$. For $\nu=1$ and $z\nu=1$, we obtain the scaling exponent $\eta_{KZM,an}=1$:
\begin{equation}
    n=\left(\frac{\tau_0}{\tau_q}\right)^1
\end{equation}
This is different from the recent prediction~\cite{Kirkby2025} for the transition from the condensed phase to the supersolid in condensates with dipolar interactions, because it was based on the parabolic dispersion relation and dipole-dipole interactions.

The analytical result obtained above for the slow-quench regime is in a good agreement with the experimental results shown in Fig.~\ref{fig3}c.  In spite of a relatively large uncertainty, the observed scaling exponent $\eta_{KZM,exp}=1.0\pm 0.2$ is clearly different from the predictions for a scalar condensate $\eta_{KZM,scalar}=1/2$ (without RDSOC or with uncoupled spin components) and from the scaling predicted numerically~\cite{Kirkby2025} for a condensate with dipolar attractive interactions  $\eta_{KZM,dip}\approx 0.3$. We can therefore claim that we have successfully observed an original Kibble-Zurek scaling in a spin-orbit coupled supersolid.

To continue our theoretical consideration, we can determine the parameters controlling the relaxation time $\tau_0$ in this expression. The dimensionless parameter defining the quench is $\varepsilon=t/\tau_q$, which enters the expression for the interaction  energy (chemical potential) as $g_1n=g\varepsilon$ ($n$ is dimensionless). Using $\Gamma=\Lambda E$, we obtain $\tau =1/(\Lambda E)=1/(\Lambda g\varepsilon)=\tau_0/|\varepsilon|^1$ and finally $\tau_0=1/(\Lambda g)$. Interestingly, it does not depend on $\alpha$.

\begin{figure}
    \centering
    \includegraphics[width=0.99\linewidth]{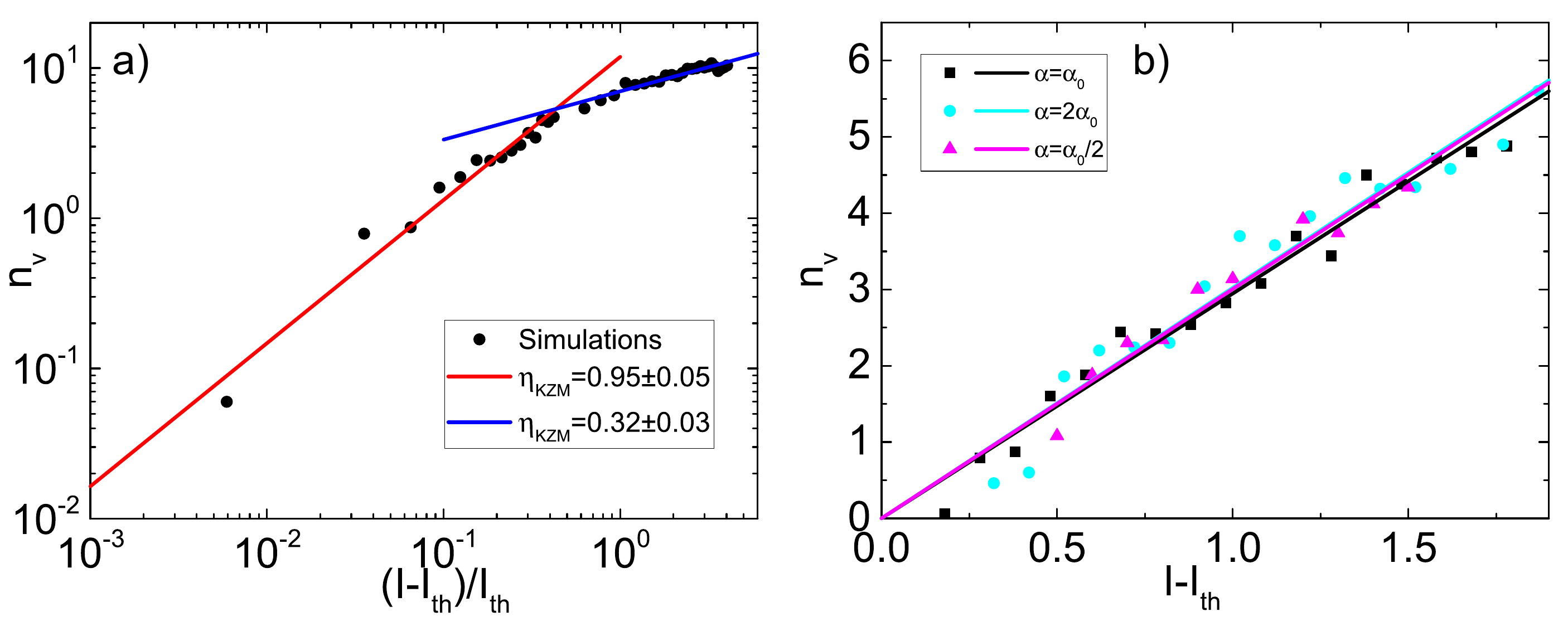}
    \caption{KZM scaling of topological defects in a polariton supersolid. a) Results of numerical simulations with slow- and fast-quench asymptotes. b) Slow-quench region for varying the RDSOC constant $\alpha$: the slope of the linear asymptote does not depend on $\alpha$}
    \label{fig2}
\end{figure}

The other limiting case $\xi\ll a$ ($k^*\gg k_0$) can be considered more qualitatively. The equality determining the scaling of the correlation length is usually written for two characteristic energies. Since we have more than two characteristic energies, it can be a combination of these quantities, for example, their average. Since we are dealing with conditional probabilities, which are multiplicative, it is relevant to use a geometric average, giving: $\sqrt{\Delta \times gn}=\sqrt{\alpha k\times  \hbar^2 k^2/(2m)}$.
Using $\mu=gn=g\varepsilon$ and rewriting this condition as an equation for the correlation length gives
\begin{equation}
    \xi\sim \frac{\xi_0}{|\varepsilon|^{1/3}}
\end{equation}
that is, $\nu=1/3$ and the resulting KZM scaling is also $\eta_{KZM,an2}=1/3$.

\textit{Numerical simulations.} We performed multiple sets of numerical simulations with different parameters based on the spinor Gross-Pitaevskii equation with relaxation and lifetime~\eqref{gpe}, in order to confirm that the agreement between experiment and theory is not accidental. For each set of parameters, we were varying the pumping intensity $\gamma_0$ (the quench parameter) above the condensation threshold. For each value of the pumping intensity, we performed a sufficient number of simulations $N$ in order to obtain a reliable estimate of the average number of vortices in the simulated (finite-size) sample: $N$ was chosen to have an approximate relative uncertainty of at most $\Delta n_V/n_V= 1.96/\sqrt{n_V N}=0.3$ where $n_v$ is the vortex number.
We take $g_2=0$, and $g_1$ such as to have a maximal value of blue shift of 1~meV above threshold. We solve Eq.~\eqref{gpe} numerically, using the parameters of the polariton dispersion and $\Lambda$ determined from the experimental results~\cite{muszynski2024observation}. We considered a square sample with a lateral size $l=128~\mu$m.

The results of numerical simulations are shown in Fig.~\ref{fig2}. The average number of vortices in our sample obtained from the simulations is shown in Fig.~\ref{fig2}(a) with black dots. A clear transition in behavior can be seen around $n_V=5$. For smaller defect densities (slow quench), the red line is a power law fit with a scaling exponent $\eta_{KZM,num}=0.95\pm 0.05$, compatible with the analytical prediction $\eta_{KZM,an}=1$. For higher densities (fast quench), the blue line is a power law fit with a scaling exponent $\eta_{KZM,num2}=0.32\pm 0.03$, also compatible with the analytical prediction $\eta_{KZM,an2}=1/3$. Figure~\ref{fig2}(b) demonstrates the slow-quench regime in linear scale for varying RDSOC constant $\alpha$: the slopes  ($3.0\pm 0.2$, $3.0\pm 0.2$ and $2.9\pm 0.2$) obtained by fitting the results of numerical simulations with a linear function show no dependence on $\alpha$, in agreement with analytical predictions (of course, the transition between the quench regimes shifts with $\alpha$). 

Comparing the experimental observations reported in Fig.~\ref{fig3}(c), the analytics and the numerical simulations shown in Fig.~\ref{fig2}(a), we observe a very good agreement between the three scaling exponents for the slow-quench regime: $\eta_{KZM,an}=1$, $\eta_{KZM,exp}=1.0\pm0.2$, and $\eta_{KZM,num}=0.95\pm 0.05$, allowing us to claim prediction and experimental observation of an original scaling exponent in a polariton spin-orbit coupled supersolid.

To conclude, we have studied the Kibble-Zurek mechanism in the formation of a supersolid via the Rashba-Dresselhaus spin-orbit coupling in a liquid crystal microcavity. We have demonstrated theoretically the existence of two different scaling laws for slow and fast quenching, appearing because of the characteristic shape of the dispersion due to spin-orbit coupling. We also present the experimental results on the statistics of topological defects in single-shot experiments with supersolid formation, confirming the theoretically predicted scaling for the slow-quench regime.

\begin{acknowledgments}
This work was supported by the European Union’s Horizon 2020 program, through a FET Open research and innovation action under the grant agreements No. 964770 (TopoLight) and EU H2020 MSCA-ITN project under grant agreement No. 956071 (AppQInfo). Additional support was provided by the ANR Labex GaNext (ANR-11-LABX-0014), the ANR program "Investissements d'Avenir" through the IDEX-ISITE initiative 16-IDEX-0001 (CAP 20-25), the ANR project MoirePlusPlus (ANR-23-CE09-0033), and the ANR project "NEWAVE" (ANR-21-CE24-0019). We are grateful to the Mésocentre Clermont-Auvergne of the Université Clermont Auvergne for providing help, computing and storage resources.
\end{acknowledgments}

\bibliography{biblio}

\renewcommand{\thefigure}{S\arabic{figure}}
\renewcommand{\theequation}{S\arabic{equation}}
\renewcommand{\thesection}{\arabic{section}}
\setcounter{figure}{0}
\setcounter{equation}{0}
\section{Supplementary}

Here, we provide additional information on the experiment and theory.

\subsection{Sample description and experiment}

The cavity consists of two distributed Bragg reflectors (DBRs) composed of 15 and 13 SiO${_2}$/Ta${_2}$O${_5}$ layers, with a central wavelength at 490~nm. The DBRs were deposited on 30~nm ITO electrode layers on quartz substrates with a flatness of $\lambda$/20(@633~nm). Both DBRs were spin-coated with 35~nm MeLPPP layers covered with protective Al$_{2}$O$_3$ layers of 20~nm thickness. Antiparallel orienting layers (SE-130, Nissan Chem., Japan) were deposited on both substrates using the spin-coating method. The wedge cavity was achieved using glass spacers with sizes of 1.5~--~2~$\mu$m placed between the substrates. The cavity was filled with a liquid crystal mixture, LC2091$^{\ast}$ (refractive indices $n_o\approx1.57$ and $n_e\approx1.98$, for the sodium line 589~nm), in the nematic phase by capillary action. We use the MeLPPP polymer as an active region ($M_n = 31,500$, $M_w = 79,000$). The strong coupling was demonstrated in~\cite{muszynski2024observation}. 

\textit{\label{sec:sm4}Experimental details.}
The LC microcavity was addressed by an AC waveform generator with 1000 Hz square signal that is phase-synchronized with the excitation laser pulses. The condensate was non-resonantly excited with pulses at 400~nm wavelength.  Coupling the 150~fs pulses into a 25~$\mu$m multi-mode fiber allows us to temporarily stretch them to several tens of ps (about 50~ps) and homogenize the beam profile and the polarization. The excitation light was focused onto the sample with a 20x objective (NA~=~0.42). The condensate emission was collected with a 10x objective (NA~=~0.30) and detected in a real-space configuration by a CCD camera. 

In order to extract the threshold intensity with high precision, we analyze the statistics of the distribution of the integrated intensity of each single shot image. We define the  threshold intensity as the part of the threshold curve (see~\cite{muszynski2024observation}) with the steepest slope, which corresponds to a minimum on the histogram of the integrated intensity distribution (steep slope means lower "density of states"). This threshold coincides with the onset of coherence and the formation of stripes in the observed single-shot spatial images.

\subsection{Four-component condensate wave functions}
The wave function describing a spinor wavefunction with two contributions corresponding to the eigenstates at two opposite wave vectors $\pm k_0$, in the absence of vortex can be written in circular basis as:
\begin{equation}
\left| \psi  \right\rangle  =\frac{1}{2} \left( {\begin{array}{*{20}{c}}
{\cos \frac{\theta }{2}}\\[2pt] 
{\sin \frac{\theta }{2}}
\end{array}} \right){e^{ - ik_0x-i\phi_1}} + \frac{1}{2}\left( {\begin{array}{*{20}{c}}
{\cos \frac{{\pi  - \theta }}{2}}\\[2pt] 
{\sin \frac{{\pi  - \theta }}{2}}
\end{array}} \right){e^{ + ik_0x-i\phi_2}}
\label{WF1}
\end{equation}
where $\theta$ is the polar angle of the Stokes vector, whereas the azimuthal angle does not explicitly appear being  fixed by the polarization H and V of the coupled modes. The phases $\phi_1$ and $\phi_2$ of the two k-components are chosen randomly due to the spontaneous symmetry breaking during the condensation process. 

Rewriting the expression allows one to make appear explicitly a random phase responsible for the position of the stripes, and another one (an overall global phase):
\begin{equation}
\left| \psi  \right\rangle  =\frac{1}{2}e^{i\zeta}\left( \left( {\begin{array}{*{20}{c}}
{\cos \frac{\theta }{2}}\\[2pt] 
{\sin \frac{\theta }{2}}
\end{array}} \right){e^{ - i(k_0x-\eta)}} + \left( {\begin{array}{*{20}{c}}
{\cos \frac{{\pi  - \theta }}{2}}\\[2pt] 
{\sin \frac{{\pi  - \theta }}{2}}
\end{array}} \right){e^{ + i(k_0x-\eta)}}\right)
\label{WF2}
\end{equation}
where the global phase is $\zeta=-(\phi_1+\phi_2)/2$, and the relative phase controlling the position of the stripes is $\eta=-(\phi_1-\phi_2)/2$.

The corresponding total density reads
\begin{equation}
\left\langle {\psi }
 \mathrel{\left | {\vphantom {\psi  \psi }}
 \right. \kern-\nulldelimiterspace}
 {\psi } \right\rangle  = \frac{1}{2} + \frac{1}{2}\cos (2(k_0 x-\eta))\sin \theta 
 \end{equation}
It exhibits oscillations, whose contrast is $U=\sin\theta$ and whose position is controlled by the randomly chosen phase $\eta$. This expression gives a direct relation between the circular (or linear) polarization degree with the contrast of the stripes observed in total intensity. 

Optical experiments allow to measure the intensity of a single linear polarization, with its amplitude obtained from the circular components as $\ket{\psi_H}=(\psi_++\psi_-)/\sqrt{2}$.

This allows to strongly improve the contrast, as required for single shot measurements:
\begin{eqnarray}
    I_H&=&\braket{\psi_H}=\sin^2\left(\frac{\pi}{4}+\frac{\theta}{2}\right)\cos^2\left(k_0 x -\eta\right)\nonumber\\
    &=&\sin^2\left(\frac{\pi}{4}+\frac{\theta}{2}\right)\left(\frac{1}{2} + \frac{1}{2}\cos 2(k_0 x-\eta)\right)
\end{eqnarray}
The contrast of this signal is always 100\% (at least without the experimental background noise), no matter the value of $\theta$, contrary to the stripes observed directly in $\braket{\psi}$, while the origin and the position of the stripes are exactly the same, governed by the phase $\eta$, as can be seen from the cosine function.

\subsection{Types of vortices}
When $\Delta=0$, there is a spin-valley locking and spin components are uniquely attributed to a given wave vector $\pm k_0$. Elementary vortices are half-vortices present only in one spin/valley components. They undergo only short range interactions if the interspin interaction constant $\alpha_2$ is nonzero which is the case we consider, being relevant to the cavity polariton case. 
When $\Delta$ is nonzero, both spin components are present in both k-valleys and the condensate really contains four components. Different type of vortices (with topological charge 1) can be in principle considered. However, for both $k_0$ and $-k_0$ components, the linear polarization of the eigenstates far from the vortex core is imposed by the Hamiltonian. It means that a vortex should embed the two spin components in a given valley which allows a fixed relative phase (governing linear polarization orientation) between the two spin components of the valley spinor. One cannot have a vortex appearing in one circular component only in a given valley, but it should be present in both with the same phase (winding). It limits the possible combinations to two vortex windings $\pm 1$ in two valleys ($k_0$ or $-k_0$), which gives four types of half-vortices. In principle, two vortices can be located at the same point in space in different valleys, potentially giving rise to full vortices.

A single vortex in $k_0$ valley means  multiplying the corresponding part of the wavefunction by $\exp(i\varphi)$, where $\varphi$ is the polar angle with respect to the vortex core. eq ~\eqref{WF1} and ~\eqref{WF2} are still valid, but replacing $\phi_{2}$ by $\phi_2+\varphi$, so that $2\eta=\phi_2-\phi_1+\varphi$. Passing from below the vortex core $(x=0,y<0)$ to above $(x=0,y>0)$ corresponds to a change of $\varphi$ by $\pi$ which corresponds to changing from a maxima of the density to a minima, which corresponds to a perfect dislocation, in the stripes density located at the vortex core. The same reasoning can be applied to the linear polarization components which also show a dislocation whereas the fringe contrast stays one. 

These vortices are also essentially half-vortices mainly associated with one of the two circular polarizations. This main circular polarization components cancels in real space, but not at the origin of the cylindrical coordinate where is the vortex core considering only one $k_0$ components. On the other hand, the other minority circular component does not show a singular point in real space.

\subsection{Kibble-Zurek scaling}
The scaling of the topological defects with the quench parameter is derived from the Landau theory of phase transitions based on the expression for the free energy density~\cite{Zurek1996}:
\begin{eqnarray}
f\left( r \right) = & -& (E_0+\mu) {\left| \psi  \right|^2} + \frac{g}{2}{\left| \psi  \right|^4} + {\psi ^*}\Delta {\sigma _x}\psi \nonumber\\
 &+& \frac{{{\hbar ^2}}}{{2m}}{\left| {\nabla \psi } \right|^2} - i\alpha {\psi ^*}{\sigma _z}\frac{\partial }{{\partial x}}\psi 
 \label{zurek}
\end{eqnarray}
Here $E_0$ sets the energy zero reference at the bottom of the dispersion, while $\mu$ is the chemical potential changing sign at the transition point.
The correlation length is determined by comparing the kinetic and interaction energy terms. We are interested in the phase fluctuations of the wave function, with the phase variation described by the wave vector $k$. In this case, the mass term and the RDSOC term give contributions proportional to $k^2$ and $k$, respectively, thus dominating at different scales. This allows to neglect the $k^2$ term during the computation in the low-$k$ limit.

\begin{figure}
    \centering
    \includegraphics[width=0.95\linewidth]{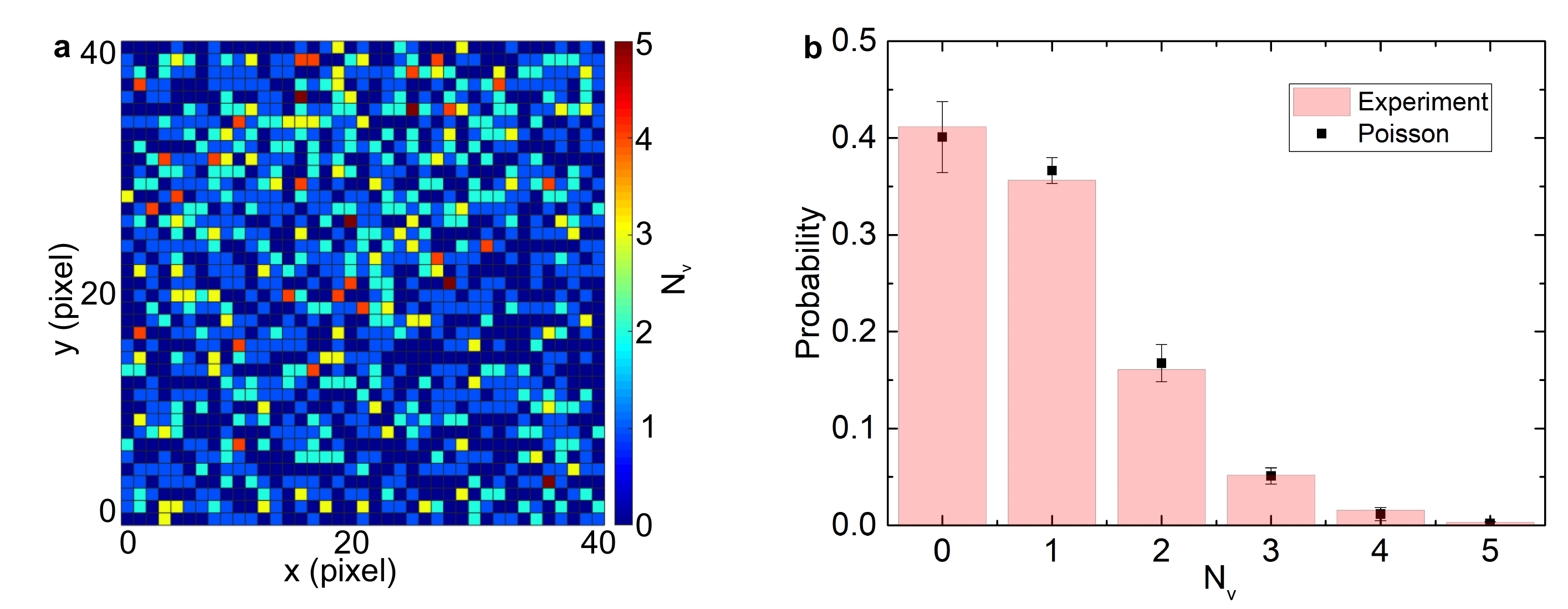}
    \caption{\textbf{Spatial distribution of vortices in experiment.} a) Number of vortices observed in each of the pixels. b) Histogram of the frequencies of the number of vortices per pixel together with the Poisson distribution and the theoretically expected uncertainty for a finite sample.
    }
    \label{EDF1}
\end{figure}

\subsection{Experimental distribution of vortices}
Figure~\ref{EDF1} shows the statistics of the spatial distribution of vortices observed in the experiment. We select a region where the intensity of the condensate is approximately constant and analyze the results of the automatic vortex detection scheme used for the observation of Kibble-Zurek scaling. Figure~\ref{EDF1}a shows a plot of the  number of vortices detected for each pixel (shown in false color). The distribution is random and does not exhibit any significant pinning, indicating the freedom of the formation of the topological defects, in agreement with the Kibble-Zurek mechanism. Figure~\ref{EDF1}b shows the histogram of the frequencies of the number of vortices for all pixels. The columns show the experimental frequencies, whereas the dots with error bars show the theoretically expected values for the Poisson distribution. The error bars correspond to the expected uncertainty of the frequencies for a finite sample of the size used in experiment. The good agreement between the theory and the experiment again indicates the randomness of the formation of the topological defects, expected for the Kibble-Zurek mechanism.

\end{document}